\documentclass[sigconf]{acmart}

\usepackage{booktabs}
\usepackage{cleveref}
\usepackage{dirtytalk}
\usepackage{natbib}
\usepackage{subfig}
\usepackage{enumitem}
\usepackage{url}
\usepackage{bbding} 
\usepackage{framed}

\setcopyright{rightsretained}

\copyrightyear{2018} 
\acmYear{2018} 
\setcopyright{acmlicensed}
\acmConference[CHASE'18]{CHASE'18: IEEE/ACM 11th International Workshop on Cooperative and Human Aspects of Software Engineering}{May 27, 2018}{Gothenburg, Sweden}
\acmPrice{15.00}
\acmDOI{10.1145/3195836.3195837}
\acmISBN{978-1-4503-5725-8/18/05}

\begin{document}

\title{On Gender, Ethnicity, and Culture in Empirical Software Engineering Research}

\author{Lucas Gren}
\orcid{1234-5678-9012}
\affiliation{%
  \institution{Chalmers University of Technology and The University of Gothenburg\\}
  \streetaddress{The Department of Computer Science and Engineering}
  \city{Gothenburg} 
  \state{Sweden} 
  \postcode{412--92}
}
\email{lucas.gren@cse.gu.se}

\begin{abstract}
This note highlights the importance of investigating diversity aspects in combination in empirical research. It draws on the psychological discourse and suggests why and how software engineering scholars can use the aspect of diversity in all their research endeavors. 
\end{abstract}

\ccsdesc[500]{General and reference~Empirical studies}
\ccsdesc[500]{General and reference~Validation}

%
%

%
%


\keywords{external validity; empirical software engineering; intersectionality}

\maketitle

\section{Introduction}\label{sec:introduction}
In recent years, empirical software engineering has gone through a similar transition as psychology in relation to explaining the differences and effects of diversity in the human population. In fact, most scientific fields that do research with human subjects have started by conducting studies in the vicinity of their home university, which is natural. Just like the field of social psychology has suffered from the assumption that the human population is white American men \citep{hogg2014sp}, most software engineering research also seems to have stemmed from research on white American or European men, and one example is who wrote the Agile Manifesto. As the field grows both in research and practice, and if the aim is to universally generalize to the part of the human population that do software engineering, a more though-through approach to external validity is needed in the field. In this short research note, I argue for why and how. 

\section{Nonrandom sampling}\label{sec:introduction}
If we want to generalize to the entire human population involved in software engineering, which we often do, the consequences of non-random sampling are severe. Researchers must understand the concept of external validity in a broader sense and editors of respected outlets should therefore impose the necessity of discussing the implications of not conducting random sampling of the intended population. Some work has been conducted in software engineering research in order to improve how generalizations are made (see e.g.\ \citet{nagappan2013diversity}), however, they do not mention the aspects of gender, ethnicity or culture explicitly in relation to research. 

Even in the much older field of social psychology, I have been quite surprised that it seems to be common practice to generalize to the human in general only sampling from white male college students. This culture in the general sciences must be changed, not primarily from an equality perspective, but from a pure scientific rigor perspective. Some studies from social psychology, like the ones presented by \citet{brescoll2010hard} and \citet{wigboldus2000we}, explicitly study diversity aspects and are therefore important first steps, but these are specific studies of inequality and stereotyping and do not, by themselves, introduce better sampling in general, if replicated in the software engineering field.

One important paper I have found instead discusses, and suggests, solutions to my defined problem of random sampling \citep{cole2009intersectionality}. In order to introduce proper random sampling of the human population in research, the author suggests making use of the following three questions through-out the research process: (1) Who is included in this category? (2) What role does inequality play? (3) Where are the similarities? The first question involves understanding diversity within social categories (see Section~\ref{sit}). The second questions conceptualizes social categories as systems of privileges and power that structure social life. The third looks for common aspects across categories often viewed as deeply different \cite{cole2009intersectionality}.

These questions try to enforce proper random sampling, but also a way to do narrower studies on groups often excluded in psychology research, in this case. \citet{cole2009intersectionality} points at the danger of using sex or race as independent variables since we lose too much resolution and attribution with such simplification. As an example the author presents the fact that Caucasian and men-of-color have different experiences of being middle-class, so race, gender, and class need to be studied in combination. Likewise, an Indian, Chinese, American, or Dutch QA engineers might have different experiences of assuring quality in a large organization in each country. 

The implications of having this diverse perspective of people is that we can properly conduct research on different organizations of different societies and get a richer understanding of the cooperative and human aspects of software engineering. As mentioned above, I believe editors should be much stricter and enforce an analysis of the threats to validity that single-category studies entail.  

\section{Social Identity Theory}\label{sit}
If we should not use independent variables to investigate social categories, how can we then find richer information regarding different social groups? One theory that has is well-researched in relation to explaining many group phenomena is the social identity theory (see e.g.\ \citet{hewstone2002intergroup,hogg2014sp}). The theory has also been shown to be valid in social neuroscience (see e.g.\ \citet{van2010social}), i.e.\ evidence has been given not only from social psychology research. The social identity is based on social categorization, which is the classification of people into different social groups. This cognitive process is a deeply rooted human trait and a person's social identity is the part of the self that is derived from the different memberships in social groups. Social identity theory is the theory of group membership and inter-group relations based on self-categorization, social comparison and a self-definition in terms of properties of all the groups an individual belongs to, and self-categorization is how we categorize ourselves and thereby construct a social identity \citep{hogg2000we}. According to the minimal group paradigm \citep{tajfel1971social}, even explicitly random group assignments trigger discriminatory behavior against groups that an individual is \emph{not} a member of. The idea is that a successful inter-group bias creates or protects (high) in-group status, which provides a positive social identity. This positive social identity, in turn, satisfies group-members' need for positive self-esteem. 

In addition, \citet{hogg2014sp} describe the \emph{looking-glass self}, which states that the \emph{self} is derived from seeing ourselves as the way we \emph{think} others see us, which can be different from how they \emph{actually} see us. This fact highlights the importance of understanding the context of different groups of people in order to understand difference and similarities in individuals. For example, a software tester that is a woman would then see herself as she thinks others see her, which then needs to be understood from investigating the dynamic context of such work. Simply testing gender differences as a binary variable just gives us a statistical significant differences in gender at best, which provides no details is what the problems are and how to solve them.

\section{Discussion}\label{sec:discussion}
In reflection to my own readings of the psychology literature in addition to my knowledge of empirical software engineering research, I have always felt dejected by the fact that there is always a statistically significant difference between men and women almost no matter the research area. Such results are initially important to realize that there are differences, but to always include, and control, this binary variable in statistical testing offers little help in dealing with the underlying issues. I believe the guidelines presented by \citet{cole2009intersectionality} could trigger richer analysis of differences and similarities of human groups, which is very much needed.

Not only is the social identity important to include in studies, but social psychology findings also offer explanation for self-categorization, which can be used for interventions in organizations. For example, the `stereotype threat' (i.e.\ the worry that one's own behavior might confirm a negative stereotype) will affect our actions and performance on different tasks at hand \citep{hogg2014sp}. Yet another example is the self-fulfilling prophecies, which is the fact that people sometimes act according to other people's expectations of them, which might not be overlapping with how they would have acted without such expectations (see e.g.\ \citet{rosenthal1994interpersonal}).

If our intended population is humans involved in software engineering, we can not base inference on only white men from America or Europe, also, empirical software engineering can not be conducted in social vacuum, meaning that better sampling or more diverse studies with regards to culture, are also needed in order to understand software engineering in general \citep{hogg2014sp}. As a closing remark, and as \citet{hogg2014sp} also write regarding psychology research, many software engineers have been white men through history and for them to start researching and making sense of their own context, is inevitable. However, the problem, again, is the aim to universally generalize all these findings without reflection. 

The time is ripe for larger and more diverse studies in empirical software engineering, which we might already see indications of with publications like e.g.\ \citet{hoda2017becoming}. My hope with this research note is to accelerate that change, and for empirical software engineering researchers to apply what is often called \emph{intersectionality} \citep{cole2009intersectionality}.


The future of, especially, global software engineering is dependent on such studies, and the importance of socio-technical coordination was defined as utterly important already in 2007 by \citet{herbsleb2007global}. In 2013, a systematic literature review by \citet{matalonga2013factors} shows that such studies still are extremely rare.

\bibliographystyle{ACM-Reference-Format}
\bibliography{references}  

\end{document}